\documentclass[12pt]{article}
\usepackage{epsfig}
\usepackage{graphicx}
\usepackage{amssymb}
\usepackage{amsmath}
\usepackage{amsfonts}
\usepackage[dvips]{color}

\newcommand{\R}{\mathbb{R}}

\newcommand{\fg}{\mathfrak{g}}

\newcommand{\fz}{\mathfrak{z}}

\newcommand{\fK}{\mathfrak{K}}

\newcommand{\cP}{\mathcal{P}}

\newcommand{\cT}{\mathcal{T}}

\newcommand{\be}{\begin{equation}}
\newcommand{\ee}{\end{equation}}
\newcommand{\bea}{\begin{eqnarray}}
\newcommand{\eea}{\end{eqnarray}}
\newcommand{\nn}{\nonumber}

\newcommand{\ed}{\end{document}}

\newcommand{\bi}{\begin{itemize}}
\newcommand{\ei}{\end{itemize}}

\newcommand{\bce}{\begin{center}}
\newcommand{\ece}{\end{center}}
\newcommand{\sgn}{\,{\rm sgn}}
\newcommand{\RE}{\,{\rm Re}}
\newcommand{\IM}{\,{\rm Im}}

\begin{document}

\title{Resonance Phenomenon Related to Spectral Singularities,
Complex Barrier Potential, and\\ Resonating Waveguides}

\author{Ali~Mostafazadeh\thanks{E-mail address:
amostafazadeh@ku.edu.tr (corresponding author)}
\\
$^*$~Department of Mathematics, Ko\c{c} University, \\ 34450
Sar{\i}yer, Istanbul, Turkey}

\date{ }
\maketitle

\begin{abstract}

A peculiar property of complex scattering potentials is the
appearance of spectral singularities. These are energy eigenvalues
for certain scattering states that similarly to resonance states
have infinite reflection and transmission coefficients. This
property reveals an interesting resonance effect with possible
applications in waveguide physics. We study the spectral
singularities of a complex barrier potential and explore their
application in designing a waveguide that functions as a resonator.
We show that for the easily accessible sizes of the waveguide and
its gain region, we can realize the spectral singularity-related
resonance phenomenon at almost every wavelength within the visible
spectrum or outside it.\vspace{2mm}

\vspace{2mm}

\end{abstract}
\vspace{5mm}


\section{Introduction}

Recently it has been noticed that by defining the Hilbert space of a
quantum system using a nonstandard inner product one can generate a
unitary time-evolution by a Hamiltonian operator that is manifestly
non-Hermitian in the standard $L^2$-inner product \cite{review}. A
typical example is $H=-\frac{d^2}{dx^2}+ix^3$. It turns out that one
can actually describe the corresponding physical system using the
standard Hilbert space and a Hermitian Hamiltonian which is much
more complicated and highly nonlocal \cite{jpa-2006a}. The
non-Hermitian Hamiltonians with the above property must have a real
spectrum and a complete set of eigenfunctions so that every wave
function can be expanded in terms of these. There are two main
mechanisms that are responsible for the incompleteness of the
eigenfunctions of a non-Hermitian Hamiltonian. These are
respectively associated with the presence of exceptional points
\cite{heiss} and spectral singularities \cite{Naimark}.

An exceptional point is a point in the space of parameters $M$ of
the Hamiltonian operator $H$ where an eigenvalue of $H$ becomes
defective, i.e., if the parameters change continuously along a path
in $M$ that passes through an exceptional point, two or more of the
eigenvalues of $H$ together with their eigenvectors coalesce.
Exceptional points have found various physical realizations and
applications \cite{eps}. They have also been the subject of
experimental studies \cite{eps-exp}. In contrast, spectral
singularities are certain points of the continuous spectrum of
non-Hermitian scattering Hamiltonians \footnote{Here we only
consider Hamiltonians of the form $H=-\frac{d^2}{dx^2}+v(x)$ where
$x$ is real and $v(x)$ is a complex scattering potential.} whose
presence ruins the completeness of the eigenfunctions, although for
each point of the spectrum (including the spectral singularity)
there corresponds two linearly independent eigenfunctions
\cite{jpa-2009}. The physical meaning and importance of this strange
mathematical phenomenon have been obscure until very recently
\cite{p89}.

The appearance of spectral singularities as an obstruction for
defining a unitary quantum system using a complex potential was
initially noted by Samsonov \cite{samsonov} who, following the
pioneering work of Naimark \cite{Naimark, Naimark-book}, only
considered systems defined on the half-line. A simple example of an
exactly solvable model defined on the whole real line that can
support a spectral singularity is the one given by a delta-function
potential, $v(x)=\fz\,\delta(x)$, with a complex coupling constant
$\fz$. As shown in \cite{jpa-2006b} this model has a spectral
singularity provided that $\fz$ is purely imaginary. The emergence
and the mechanism by which a spectral singularity spoils the
completeness of the eigenfunctions of a scattering non-Hermitian
Hamiltonian are studied in \cite{jpa-2009}.

In Ref.~\cite{p89} we have provided a physical interpretation for
the spectral singularities that identifies them with the energies of
scattering states having infinite reflection and transmission
coefficients. Because this is a characteristic property of resonance
states, spectral singularities correspond to the resonance states
having a real energy and hence a zero width. The existence of such
states reveals a new type of resonance effect with potential
applications in various areas of physics. To demonstrate how one can
realize this effect, we have explored, in Ref.~\cite{p89}, the
occurrence of spectral singularities for the $\cP\cT$-symmetric
imaginary barrier potential:
    \be
    v(x):=\left\{\begin{array}{ccc}
    -i\zeta\sgn(x) & {\rm for} & |x|<\alpha,\\
    0& {\rm for} & |x|\geq \alpha,\end{array}\right.
    \label{v-PT=}
    \ee
where $\zeta$ and $\alpha$ are real parameters, $\alpha$ is
positive, and $\sgn(x)$ stands for the sign of $x$
\footnote{$\sgn(x):=x/|x|$ if $x\neq 0$, and $\sgn(0):=0$.}.
\begin{figure}
\begin{center}
\includegraphics[scale=.85,clip]{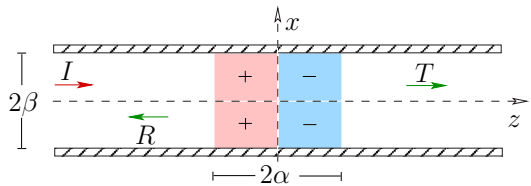}
\caption{(Color online) Cross section of a rectangular waveguide
with gain~(+) and loss~(-) regions in the $x$-$z$ plane. Arrows
labeled by $I$, $R$, and $T$ represent the incident, reflected, and
transmitted waves.\label{fig0}}
\end{center}
\end{figure}
The potential (\ref{v-PT=}) can be used to model the propagation of
certain transverse electric fields in a planar slab waveguide with
an adjacent pair of loss and gain regions \cite{rdm} as shown in
Figure~\ref{fig0}. In \cite{p89}, we have determined the values of
physical parameters of this system that correspond to a spectral
singularity. It turns out that for fixed values of the parameters
entering the permittivity, spectral singularities occur for a
discrete set of values for the length of the loss and gain regions
$2\alpha$ and the height of the waveguide $2\beta$. For the
fundamental mode of the standing wave formed in the transverse
direction (along the $x$-axis), the largest allowed values of
$\alpha$ and $\beta$ are respectively $1004.17$~nm and $62.0464$~nm.
Explicit calculation shows that for these value of $\alpha$ and
$\beta$, the reflection and transmission coefficients (at the
resonance frequency) exceed $10^{14}$. Note however that this
requires maintaining this value of $\alpha$ for the length of the
loss and gain regions to an extremely high degree of precision
($10^{-1}$ Angstroms). A 0.1\% deviation from the above values of
$\alpha$ and $\beta$ leads to a decrease in the reflection and
transmission coefficients by a factor of $10^7$. Furthermore, the
experimental realization of spectral singularities that is proposed
in \cite{p89} involves restricting the frequency of propagating wave
to the resonance frequency of the active region and demands that the
imaginary part of the permittivity for the loss and gain differ only
by a sign.

In the present article, we study the spectral singularities of the
complex barrier potential
    \be
    v(x):=\left\{\begin{array}{ccc}
    \fz & {\rm for} & |x|<\alpha,\\
    0& {\rm for} & |x|\geq \alpha,\end{array}\right.
    \label{v=}
    \ee
where $\fz$ is an arbitrary complex coupling constant. Our main
motivation is that the proposed experimental setup that would
confirm the resonance phenomenon associated with the spectral
singularities of (\ref{v=}) is free of most of the above-mentioned
restrictions. This is essentially because of the additional degree
of freedom, namely the real part of $\fz$ in (\ref{v=}). Another
reason for considering the potential (\ref{v=}) is that, similarly
to (\ref{v-PT=}), the mathematical problem of locating the spectral
singularities of (\ref{v=}) admits an exact and essentially analytic
solution. This is very convenient, for it enables us to avoid
dealing with the subtleties of the numerical treatments of the
problem.

A most remarkable result of our investigation is that for the
experimentally easily accessible dimensions of the waveguide
associated with (\ref{v=}), we can produce the spectral
singularity-related resonance effect at almost every wavelength
within the visible spectrum or outside it. This is of great
practical significance, for it may form the basis of a new mechanism
for producing laser beams of desired wave length.

This article is organized as follows. In Section~2, we provide a
brief review of the results of \cite{jpa-2009,p89} on the
calculation of spectral singularities and their relation to the
reflection and transmission coefficients. In Section~3, we solve the
problem of locating the spectral singularities of the complex
barrier potential. In Section~4, we describe a waveguide modeled
using this potential and explore the possibility of the detection of
the resonance effect associated with the spectral singularities. In
Section~5, we present our concluding remarks.

\section{Spectral Singularities}

Consider the Hamiltonian operator $H=-\frac{d^2}{dx^2}+v(x)$ for a
complex scattering potential $v(x)$ with $x\in\R$. Suppose that
$v(x)$ tends to zero as $x\to\pm\infty$ so rapidly that the integral
$\int_{-\infty}^\infty (1+|x|)|v(x)|dx$ converges. Then the spectrum
of $H$ has a real continuous part. Suppose for simplicity that the
spectrum coincides with $[0,\infty)$. Then the eigenvalue equation
$H\psi=E\psi$ yields the following asymptotic expressions for the
(generalized) eigenfunctions of $H$.
    \be
    \psi_k^{\fg}(x)\to A_\pm^\fg e^{ikx}+B_\pm^\fg
    e^{-ikx},~~~~{\rm for}~~~~x\to\pm\infty.
    \label{asmp}
    \ee
Here $\fg$ is a degeneracy label taking values 1 and 2,
$k:=\sqrt{E}$, and $A_\pm^\fg$, $B_\pm^\fg$ are possibly
$k$-dependent complex coefficients that are related by the so-called
transfer matrix $\mathbf{M}=(M_{ij})$ according to
    \be
    \left(\begin{array}{c} A_+^\fg\\B_+^\fg\end{array}\right)=\mathbf{M}
    \left(\begin{array}{c} A_-^\fg\\B_-^\fg\end{array}\right).
    \label{transfer}
    \ee
The transfer matrix encodes all the necessary information about the
scattering properties of the system. It is easy to show that it has
a unit determinant \cite{jpa-2009}.

Next, consider the Jost solutions $\psi_{k\pm}$ of the eigenvalue
equation $H\psi=k^2\psi$ that by definition satisfy
    \be
    \psi_{k\pm}(x)\to e^{\pm ikx}~~~~{\rm for}~~~~x\to\pm\infty.
    \label{jost}
    \ee
Denoting the coefficients $A_\pm^\fg$ and $B_\pm^\fg$ for the Jost
solutions $\psi_{k\pm}$ by $A^\pm_\pm$ and $B^\pm_\pm$ and using of
(\ref{asmp}) -- (\ref{jost}), we find $A_+^+=B_-^-=1$,
$A_-^-=B_+^+=0$, $A_-^+=B_+^-=M_{22}$, $A_+^-=M_{12}$, and
$B_-^+=-M_{21}$, \cite{jpa-2009}. These relations identify
$\psi_{k\pm}$ with the left- and right-going scattering solutions
\cite{muga}. As a result, the left and right transmission $T^{l,r}$
and reflections $R^{l,r}$ amplitudes are given by \cite{p89}
    \be
    T^l=T^r=\frac{1}{M_{22}},~~~R^l=-\frac{M_{21}}{M_{22}},~~~
    R^r=\frac{M_{12}}{M_{22}}.
    \label{amplitudes}
    \ee

Spectral singularities of $H$ are eigenvalues $E=k^2$ for which
$\psi_{k+}$ and  $\psi_{k-}$ become linearly-dependent
\cite{Naimark,ss-math,guseinov}. They are given by the real zeros of
$M_{22}$ where both the transmission and reflections coefficients
diverge \cite{p89}. Because the latter is a characteristic feature
of resonances \cite{siegert}, we associate spectral singularities
with a peculiar type of resonances which satisfy the eigenvalue
equation $H\psi=k^2\psi$ for a real $k^2$. These are resonances
having a zero width, \cite{p89}.

\section{Complex Barrier Potential}

Consider the complex barrier potential~(\ref{v=}). The solution of
the time-independent Schr\"odinger equation, $H\psi=k^2\psi$, for
this potential is elementary. We can use this solution, to determine
the entries of the transfer matrix $\mathbf{M}$:
    \bea
    M_{11}&=&\frac{e^{-2i\alpha k}f(w,-\alpha k)}{4w},\\
    M_{12}&=&-M_{21}=
    \frac{i(w^2-1)\sin(2 \alpha k w)}{2w},\\
    M_{22}&=&\frac{e^{2i\alpha k}f(w,\alpha k)}{4w},
    \label{Ms=}
    \eea
where $w:=\sqrt{1-\fz/k^2}$ \footnote{We use $\sqrt{1-\fz/k^2}$ to
denote the principal value of $(1-\fz/k^2)^{1/2}$ whose argument
belongs to $[0,\pi)$.}, and for all $\chi\in\R$,
    \be
    f(w,\chi):=e^{-2i\chi w}(1+w)^2-e^{2i\chi w}(1-w)^2.
    \label{f=}
    \ee
According to (\ref{Ms=}), spectral singularities are associated with
real values of $k$ for which
    \be
    f(w,\alpha k)=f(\sqrt{1-\fz/k^2},\alpha k)=0.
    \label{f=0}
    \ee
This is a complex transcendental equation involving four real
variables $\alpha, k, \RE(\fz)$ and $\IM(\fz)$, where ``Re'' and
``Im'' stand for the real and imaginary part of their argument. A
direct numerical solution of (\ref{f=0}) would amount to identifying
two of the variables as parameters (say $\RE(\fz)$ and $\IM(\fz)$)
and solving (\ref{f=0}) for the remaining two variables (i.e.,
$\alpha$ and $k$). This requires selecting a range of values for the
parameters and performing a numerical solution of (\ref{f=0}) for
all the values of the parameters within this range. This is not a
straightforward procedure, for in principle there is a real solution
only for a one-dimensional subset of the two-dimensional parameter
space. Numerical treatments may miss part of this one-dimensional
subset and could lead to various types of errors \footnote{For
example typical numerical shemes will produce $k$ values with a
small imaginary part. These do not correspond to scattering states
and must therefore be rejected.}. In the following, we shall avoid
these difficulties by offering an exact and essentially analytic
solution of (\ref{f=0}). Specifically, we shall reduce (\ref{f=0})
to a single real transcendental equation involving two real
variables and a discrete label. As we will see this provides a much
more efficient and precise method of locating spectral singularities
of the potential (\ref{v=}).

We begin our analysis by recalling that in light of (\ref{f=}), we
can express (\ref{f=0}) in the form
    \be
    e^{i\alpha k w}(1-w)=\pm e^{-i\alpha k w}(1+w).
    \ee
This is equivalent to
    \be
    \cos(2\alpha k\sqrt{1-\fz/k^2})=\pm\left(\frac{2k^2}{\fz}-1\right).
    \label{cos=rat}
    \ee
We can reduce (\ref{cos=rat}) into a pair of real equations. To this
end, we first introduce
    \bea
    \rho&:=&\frac{\RE(\fz)}{k^2},~~~~\sigma:=\frac{\IM(\fz)}{k^2},~~~~
    y:=\frac{\sigma}{1-\rho},
    \label{y}\\
    q&:=&\alpha k\sqrt{2|1-\rho|(\sqrt{y^2+1}-1)}\,\sgn(y),
    \label{q}\\
    r&:=&\alpha k\sqrt{2|1-\rho|(\sqrt{y^2+1}+1)}.
    \label{r}
    \eea
In light of (\ref{q}), (\ref{r}), and $k>0$, we have
    \be
    \sgn(q)=\sgn(y)~~~{\rm and}~~~r>0.
    \label{signs}
    \ee
Next, by inserting $\fz/k^2=\rho+i\sigma$ into (\ref{cos=rat}) and
using the elementary properties of the trigonometric and hyperbolic
functions, we express the real and imaginary parts of this equation
as
    \bea
    \cos r\:\cosh
    q&=&\pm\frac{1-(1-\rho)^2(y^2+1)}{(1-\rho)^2y^2+\rho^2},
    \label{e1}\\
    \sin r\:\sinh q&=&\mp\frac{2(1-\rho)y}{(1-\rho)^2y^2+\rho^2}.
    \label{e2}
    \eea
Here either the top or the bottom sign should be taken in both of
the equations.

We can easily eliminate $r$ in (\ref{e1}) and (\ref{e2}). We do this
by solving for $\cos r$ and $\sin r$ and imposing the identity
$\sin^2r+\cos^2r=1$. Expressing $\cosh^2 q$ that appears in the
resulting equation as $1+\sinh^2q$, we find after some lucky
cancelations
    {\small\[\sinh^4q-\left(\frac{4(1-\rho)}{(1-\rho)^2y^2+\rho^2}\right)\sinh^2q-
    \left(\frac{4(1-\rho)^2y^2}{[(1-\rho)^2y^2+\rho^2]^2}\right)=0.\]}%
Solving this equation for $q$ and using (\ref{signs}), we obtain
    \be
    q=\sgn(y) Q(\rho,y),
    \label{q=Q1}
    \ee
where
    \be
    Q(\rho,y):=\sinh^{-1}\left[\sqrt{\frac{2(\sqrt{y^2+1}|1-\rho|+1-\rho)}{
    (1-\rho)^2y^2+\rho^2}}\right].
    \label{q1}
    \ee

Next, we substitute (\ref{q=Q1}) in (\ref{e1}) and solve for $r$.
Employing the identity $\cosh(\sinh^{-1}(x))=\sqrt{1+x^2}$ to
simplify the resulting expression, we then find
    \be
    r= R_{n}^\epsilon(\rho,y),
    \label{r=R}
    \ee
where $n$ is an integer, $\epsilon$ is a sign, and
    \be
    R_{n}^\pm(\rho,y):=\pi n\pm\cos^{-1}\left(\frac{1-|1-\rho|\sqrt{y^2+1}}{
    \sqrt{(1-\rho)^2y^2+\rho^2}}\right).
    \label{R}
    \ee
Three remarks are in order.
\begin{enumerate}
\item Because for real values of $\rho$ and $y$, we have
    \bea
    &&|1-\rho|\sqrt{y^2+1}+\rho-1\geq 0,\\
    &&(1-\rho)^2y^2+\rho^2=\left(1-|1-\rho|\sqrt{y^2+1}\right)^2+
    2\left(|1-\rho|\sqrt{y^2+1}+\rho-1\right),~~~~~
    \eea
the argument of $\cos^{-1}$ in (\ref{R}) lies in the interval
$[-1,1]$. This implies that the second term on the right-hand side
of (\ref{R}) takes values in $[0,\pi]$, and $R_{n}^\pm(\rho,y)$ is
always real. Furthermore, $\sgn(R_{n}^\pm(\rho,y))=\sgn(n)$ for
$n\neq 0$, and $\sgn(R_{0}^\pm(\rho,y))=\pm$ except for the case
$\rho=y=0$ which corresponds to a real barrier potential. Therefore,
to ensure that $r>0$, we must demand that either $n>0$ and
$\epsilon$ be arbitrary or $n=0$ and $\epsilon=+$. In the following
we shall confine our attention to these two cases.

\item Every $r$ and $q$ that satisfy (\ref{e1}) and
(\ref{e2}) must also satisfy (\ref{q=Q1}) and (\ref{r=R}), but the
converse is not true. It turns out that depending on the choice of
the top or bottom sign in (\ref{e1}) and (\ref{e2}) and range of
values of $\rho$ and $y$ one must further restrict the values of $n$
and $\epsilon$ in (\ref{r=R}).

\item Equations (\ref{e1}) and (\ref{e2}) are not equivalent to
(\ref{cos=rat}) unless we impose (\ref{q}) and (\ref{r}). First, we
impose (\ref{r}). This allows us to express $\alpha k$ in terms of
$r,\rho$ and $y$. In view of (\ref{r=R}), we then find
    \be
    \alpha k=G_n^\epsilon(\rho,y):=
    \frac{R_{n}^\epsilon(\rho,y)}{\sqrt{2|1-\rho|(\sqrt{y^2+1}+1)}}.
    \label{gamma=}
    \ee
Unlike (\ref{r}) that determines $\alpha k$ in terms of $\rho$ and
$y$, (\ref{q}) gives rise to an additional constraint on the
possible values of $\rho$ and $y$. We will next derive an explicit
form of this constraint.
\end{enumerate}

We can use (\ref{q}) and (\ref{r}) to obtain
    \be
    q=r\sqrt{\frac{\sqrt{y^2+1}-1}{\sqrt{y^2+1}+1}}\:\sgn(y).
    \label{q-r}
    \ee
If we insert (\ref{r=R}) in this equation and use (\ref{signs}), we
arrive at
    \be
    q=\sgn(y) \tilde Q^\epsilon_{n}(\rho,y)
    \label{q=Q2}
    \ee
where
    \be
    \tilde Q^\epsilon_{n}(\rho,y):=R_{n}^\epsilon(\rho,y)
    \sqrt{\frac{\sqrt{y^2+1}-1}{\sqrt{y^2+1}+1}}.
    \label{Q2}
    \ee
Recall that here $n$ is a nonnegative integer, and for $n=0$ we have
$\epsilon=+$.

Enforcing (\ref{q=Q1}) and (\ref{q=Q2}) yields $Q(\rho,y)=\tilde
Q^\epsilon_{n}(\rho,y)$. This gives an infinite sequence of
equations that determine the locus of points in the $\rho$-$y$ plane
for which the complex barrier potential has a spectral singularity.
These form an infinite sequence of curves. Clearly,
$Q(\rho,y)=\tilde Q^\epsilon_{n}(\rho,y)$ is equivalent to
$F_n^\epsilon(\rho,y)=0$ where
    \be
    F_n^\epsilon(\rho,y):=\frac{1}{2}\left[\sinh^2Q(\rho,y)-
    \sinh^2\tilde Q^\epsilon_n(\rho,y)\right].
    \label{F}
    \ee
In view of (\ref{q1}), (\ref{Q2}), and (\ref{F}) we have the
following explicit expression for $F_n^\epsilon$.
    {\bea
    F_n^\pm(\rho,y)&=&
    \frac{|1-\rho|\sqrt{y^2+1}+1-\rho}{(1-\rho)^2y^2+\rho^2}\nn\\
    &&-\frac{1}{2}\sinh^2\left\{\sqrt{\frac{\sqrt{y^2+1}-1}{\sqrt{y^2+1}+1}}
    \left[\pi n\pm\cos^{-1}\left(\frac{1-|1-\rho|\sqrt{y^2+1}}{
    \sqrt{(1-\rho)^2y^2+\rho^2}}\right)\right]
    \right\}.~~~~~~~
    \label{eq}
    \eea}%
In the $\rho$-$\sigma$ plane, which coincides with the complex
$\fz/k^2$-plane, the spectral singularities are located on the
curves
    \be
    C_n^\epsilon:=\left\{(\rho,\sigma)\in\R^2\,\left| \,
    F_n^\epsilon(\rho,\mbox{$\frac{\sigma}{1-\rho}$})=0\right.\,\right\}.
    \label{extra-eqn1}
    \ee
As we will see below, not all the points on these curves turn out to
correspond to a spectral singularity.

\begin{figure}
\begin{center} 
\includegraphics[scale=.075,clip]{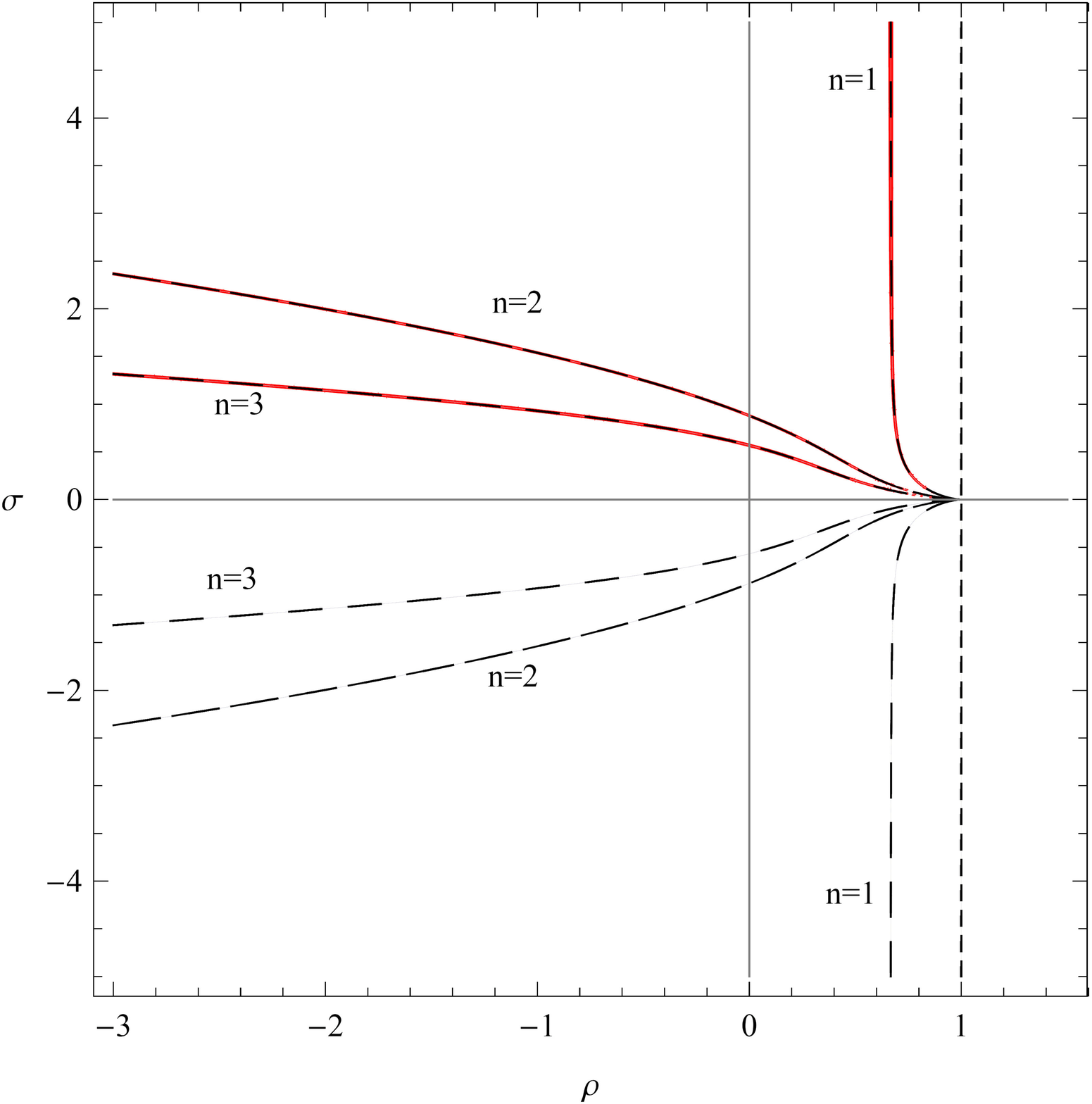}
{\caption{(Color online) Graphs of the curves $C^-_n$ (the dashed
black curves) and $\tilde C^-_n$ (the solid red curves) in the
$\rho$-$\sigma$ plane for $n=1,2,3$. $C^-_n$ are symmetric about the
$\rho$-axis. $\tilde C^-_n$ coincides with $C^-_n$ for $\sigma>0$.
This shows the lack of spectral singularities for $\sigma\leq 0$.
The dashed black line corresponds to $\rho=1$. The solid gray lines
are the coordinate axes. \label{fig1}}}
\end{center}
\end{figure}

Examining the graphs of $C_n^\epsilon$, see Figure~\ref{fig1}, we
find that these curves all lie to the left of the line: $\rho=1$ in
the $\rho$-$\sigma$ plane. The curve $C_0^+$ intersects this line at
$(\rho=1,\sigma=0)$ but this point does not correspond to a spectral
singularity, because for $\sigma=0$ the potential is real. These
observations allow us to confine our attention to the case $\rho<1$.
A consequence of this relation and (\ref{signs}) is
    \be
    \sgn(q)=\sgn(y)=\sgn(\sigma)=\sgn(\IM(\fz)).
    \label{sign-z}
    \ee

In order to determine the choices of $n$ and $\epsilon$ for which
$C^\epsilon_n$ includes points corresponding to spectral
singularities, we proceed as follows. First, we recall that the
spectral singularities are give by (\ref{f=0}). Next, we use the
identities $w=\sqrt{1-\rho-i\sigma}=\sqrt{1-\rho-i(1-\rho)y}$,  and
Eqs.~(\ref{f=}) and (\ref{gamma=}) to express the left-hand side of
(\ref{f=0}) in terms of $\rho$ and $y$. This yields a sequence of
equations of the form $\tilde F^\epsilon_n(\rho,y)=0$, where
    {\small \bea
    \tilde F^\epsilon_n(\rho,y)&:=&
    e^{-2iG_n^\epsilon(\rho,y)
    \sqrt{1-\rho-i(1-\rho)y}}\left[1+\sqrt{1-\rho-i(1-\rho)y}\right]^2
    \nn\\
    &&-e^{2iG_n^\epsilon(\rho,y)
    \sqrt{1-\rho-i(1-\rho)y}}\left[1-\sqrt{1-\rho-i(1-\rho)y}\right]^2\!\!,
    \nn
    \eea}%
and $G^\epsilon_n$ is introduced in (\ref{gamma=}). The spectral
singularities lie on the curves
    \be
    \tilde C^-_n :=\left\{(\rho,\sigma)\in\R^2\,\left| \,
    |\tilde F^\epsilon_n(\rho,\mbox{$\frac{\sigma}{1-\rho}$})|=0
    \right.\,\right\}.
    \label{tilde-C-def}
    \ee
It turns out that $|\tilde
F^\epsilon_n(\rho,\mbox{$\frac{\sigma}{1-\rho}$})|=0$ only for
$\epsilon=-$ and $n>0$. Furthermore, as shown in Figure~\ref{fig1},
the graph of $\tilde C^-_n$ coincides with a part of the graph of
$C^-_n$ that lies above the $\rho$-axis in the $\rho$-$\sigma$
plane. That is
    \be
    \tilde C^-_n =\left\{(\rho,\sigma)\in\R^2\,\left| \,
    F_n^-(\rho,\mbox{$\frac{\sigma}{1-\rho}$})=0,~
    \sigma>0\right.\,\right\}.
    \label{tilde-C=}
    \ee
This provides a highly nontrivial check on the validity of our
calculations. It also reveals the interesting fact that spectral
singularities can appear only for the cases that the coupling
constant $\fz$ has a positive imaginary part. In Section~4, we will
provide a simple physical justification for this fact.

\begin{figure}[t]
\begin{center} 
\includegraphics[scale=.075,clip]{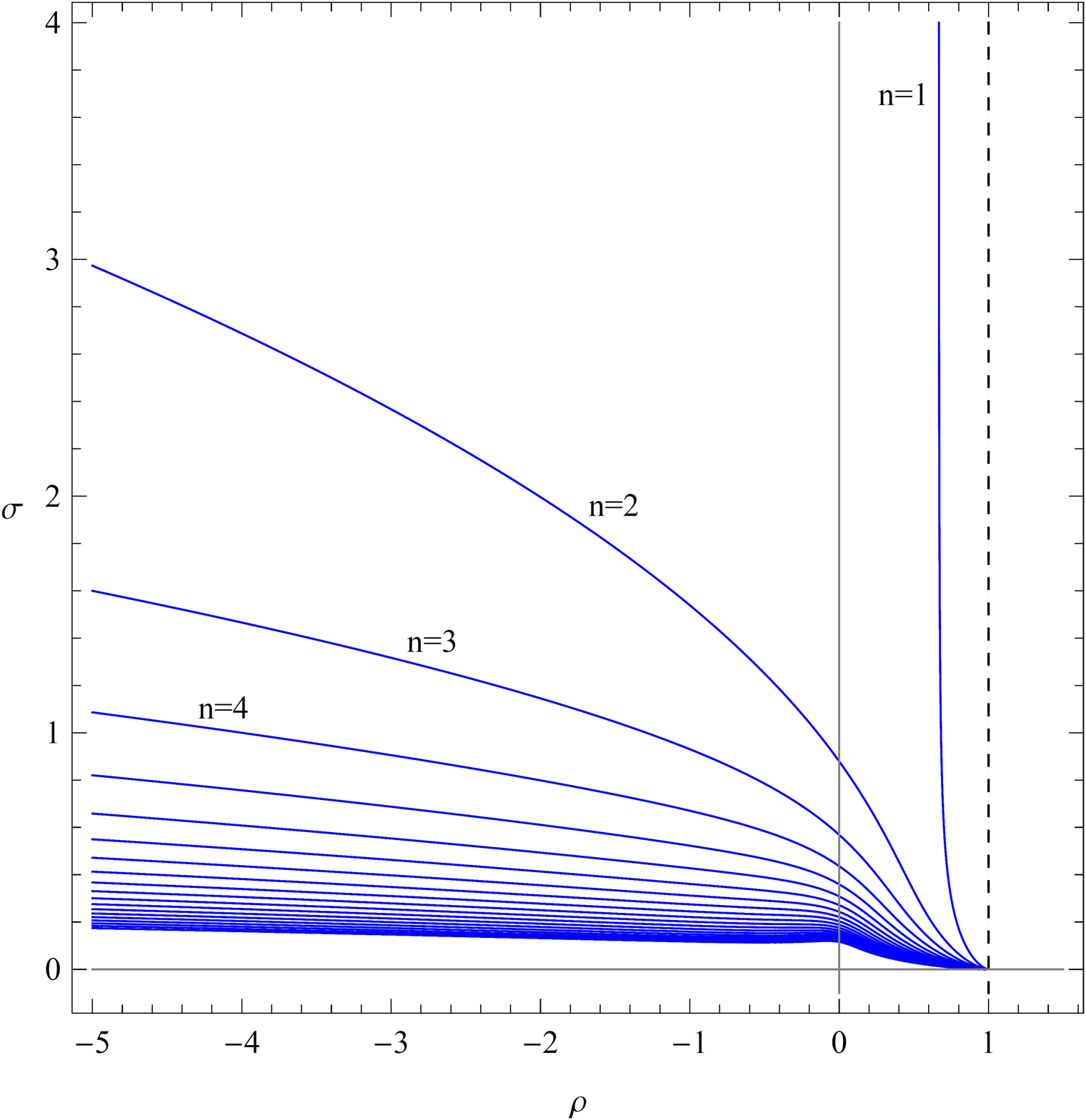}
{\caption{(Color online) Graphs of the curves $\tilde C^-_n$ in the
$\rho$-$\sigma$ for $1\leq n\leq 20$. The unmarked curves correspond
to $n=5,6,7,\cdots,20$ from top to bottom, respectively. $\tilde
C^-_1$ has an asymptote given by $\rho\approx 0.667$. The dashed
line corresponds to $\rho=1$. The solid gray lines are the
coordinate axes. \label{fig2}}}
\end{center}
\end{figure}

Another outcome of examining the curves $\tilde C^-_n$ is that, in
comparison to (\ref{tilde-C=}), the use of (\ref{tilde-C-def}) in
ploting $\tilde C^-_n$ leads to substantially larger numerical
errors. Therefore, in practice, it is more appropriate to employ
(\ref{tilde-C=}) to locate the spectral singularities of the
potential (\ref{v=}) graphically. For this reason, in preparing
Figure~\ref{fig2}, we have used (\ref{tilde-C=}) to plot $\tilde
C^-_n$ for $1\leq n\leq 20$.
This figure reveals the following facts about the curves of spectral
singularities.
    \begin{itemize}
    \item  $\tilde C^-_1$ has a vertical asymptote
    given by $\rho\approx 0.667$.
    \item For all $n\geq 1$, $\tilde C^-_n$ is a
    decreasing curve.
    \item For $n\geq 2$ the graph of $\tilde C^-_{n}$
    lies above that of $\tilde C^-_{n+1}$.
    \item As $n\to\infty$, $\tilde C^-_n$
    tends to the $\rho$-axis,
    \footnote{This can be easily shown using (\ref{eq})).}.
    \end{itemize}

\section{Resonance Effect Associated with Spectral Singularities}

In this section, we describe an electromagnetic waveguide that can
be used to realize the resonance effect associated with the spectral
singularities of the complex barrier potential~(\ref{v=}).

\begin{figure}
\begin{center} 
\includegraphics[scale=.65,clip]{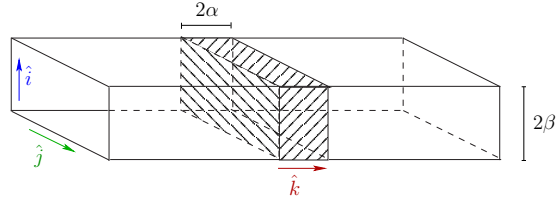}
{\caption{(Color online) A rectangular waveguide aligned along the
$z$-axis. The dashed region is filled with an atomic gas. $\hat i$,
$\hat j$, $\hat k$ are the unit vectors along the $x$-, $y$-, and
$z$-axes, respectively. \label{fig3}}}
\end{center}
\end{figure}

Let $\alpha,\beta,\gamma$ be positive parameters having the
dimension of length. Consider a rectangular waveguide with perfectly
conducting walls that guides transverse electric (TE) waves along
the $z$-axis in the region defined by $|x|<\beta$ and $|y|<\gamma$.
Suppose that the region $|z|<\alpha$ inside the waveguide is filled
with an atomic gas so that the relative permittivity inside the
waveguide is given by \cite{jackson,rdm}
    \be
    \varepsilon(z)=\left\{\begin{array}{ccc}
    1-\mbox{$\displaystyle\frac{\omega^2_p}{
    \omega^2-\omega_0^2+2i\delta\omega}$}&{\rm for}& |z|<\alpha,\\
    1&{\rm for}& |z|\geq\alpha,\end{array}
    \right.
    \label{perm}
    \ee
where $\omega,\omega_p,\omega_0$ and $\delta$ are respectively the
frequency of the wave, the plasma frequency, the resonance
frequency, and the damping constant. Figure~\ref{fig3} provides a
schematic illustration of the waveguide.

Let $\fK:=\omega/c$ be the wave number, $\fK_m:=\pi m/(2\beta)$ for
every positive integer $m$, and $\phi(z)$ be a solution of
    \be
    \phi''(z)+[\fK^2\varepsilon(z)-\fK_m^2]\phi(z)=0,
    \label{e3n}
    \ee
with a continuous derivative. Then it is an easy exercise to show
that the following TE wave is a solution of the Maxwell equations
with appropriate boundary conditions for the above waveguide
\cite{p89}.
    {\small \bea
    \vec E(\vec r,t)&=&{\cal E}~\RE \left\{e^{-i\omega t}\Big(-i\omega
    \sin[\fK_m(x+\beta)]\,\phi(z)\Big)\hat j\right\},~~~~
    \label{e1n}\\
    \vec B(\vec r,t)&=&{\cal E}~\RE \left\{e^{-i\omega t}\Big(
    \sin[\fK_m(x+\beta)]\,\phi'(z)\hat i-
    \fK_m\cos[\fK_m(x+\beta)]\,\phi(z)\hat k\Big)\right\}.
    \label{e2n}
    \eea}%
Here ${\cal E}$ is a real constant and $\hat i$, $\hat j$, $\hat k$
are respectively the unit vectors along the $x$-, $y$-, and $z$-axes
\footnote{Note that we use physicists' convention of expressing the
time-harmonic part of the wave as $e^{-i\omega t}$ as opposed to
engineers' convention that uses $e^{i\omega t}$. This leads to a
change of sign in the imaginary part of permittivity (and complex
refractive index.)}.

Eq.~(\ref{e3n}) coincides with the eigenvalue equation,
    \[\left[-\frac{d^2}{dz^2}+v(z)\right]\phi=k^2\psi,\]
for the complex barrier potential~(\ref{v=}), if we replace $x$ with
$z$ in (\ref{v=}) and make the following identifications.
    \bea
    &&k=\sqrt{\fK^2-\fK_m^2}=
    \frac{\omega}{c}\sqrt{1-\frac{\Omega^2}{\omega^2}},
    ~~~~\Omega:=\frac{\pi mc}{2\beta},~~~~
    \label{e4a}\\
    &&\RE(\fz)=\frac{\omega^2\omega_p^2(\omega^2-\omega_0^2)}{
    c^2[(\omega^2-\omega_0^2)^2+4\omega^2\delta^2]},\\
    &&\IM(\fz)=\frac{-2\omega^3\omega_p^2\delta}{c^2[(\omega^2-\omega_0^2)^2+
    4\omega^2\delta^2]}.
    \label{e4}
    \eea
These imply
    \be
    \begin{array}{c}
    \rho=\mbox{ $\displaystyle\frac{\omega_p^2(\omega^2-\omega_0^2)}{
    [(\omega^2-\omega_0^2)^2+4\omega^2\delta^2]
    (1- \Omega^2/\omega^2)}$},\\ \\
    \sigma=\mbox{ $\displaystyle\frac{-2\omega\omega_p^2\delta}{[(\omega^2-\omega_0^2)^2+
    4\omega^2\delta^2](1- \Omega^2/\omega^2)}$}.\end{array}
    \label{e6}
    \ee
For a TE wave given by (\ref{e1n}) and (\ref{e2n}) that propagates
in the waveguide, we have $\fK>\fK_m$, so that $k$ is real and
positive. Therefore, we can use the results of Section~3 to
determine the reflection and transmission amplitudes of the wave. In
particular, if we can arrange the physical parameters of the system
namely $\omega,\omega_0,\omega_p,\delta,\alpha$ and $\beta$ so that
$k^2$ is a spectral singularity, reflection and transmission
coefficients diverge and the waveguide functions as a resonator.

For the system we described above $\omega_p^2$ and $\delta$ are
positive. In view of (\ref{e4}) this implies that $\IM(\fz)<0$.
Therefore, as we showed in Section~3 (Figure~\ref{fig1}), there is
no spectral singularities. This is actually to be expected, because
for the physical system that we consider the presence of a spectral
singularity means the emission of highly amplified electromagnetic
waves. This is clearly forbidden by the law of conservation of
energy. It is quite remarkable that the absence of spectral
singularities for the barrier potentials with $\IM(\fz)\leq 0$,
which is a purely mathematical result, can be interpreted as a
direct consequence of conservation of energy in classical
electrodynamics.

We can avoid the above conflict with energy conservation, if we
consider the situation that the sign of $\omega_p^2$ is reversed. In
this case the atomic gas confined in the region $|z|<\alpha$ inside
the waveguide acts as a gain medium \cite{yariv-yeh}. We can achieve
this by shining a laser beam along the $y$-axis to excite the
resonant atoms and induce a population inversion
\footnote{$\omega_p^2$ is proportional to the difference between the
number density of atoms in the ground and excited states which
becomes negative due to population inversion \cite{yariv-yeh}.}.
This was also the mechanism we employed in \cite{p89} to realize the
resonance effect related with the spectral singularities of the
$\cP\cT$-symmetric barrier potential (\ref{v-PT=}). The main
advantage of the system we consider here is that, for the cases that
$\omega_0,\omega_p$, and $\delta$ are fixed, the spectral
singularities occur along an infinite set of curves in the relevant
parameter space, whereas for the system considered in \cite{p89}
they occur at an infinite set of isolated points.

Another practical shortcoming of the system studied in \cite{p89} is
that it relied on the assumption that the frequency of the
propagating wave $\omega$ is equal to the resonance frequency of the
gain medium $\omega_0$. Here we do not need to make such an
assumption. This provides an additional flexibility in our attempt
to adjust the parameters of the system so that the frequency of the
propagating wave coincides with that of a spectral singularity. To
demonstrate how we can do this, we first choose some typical values
for $\omega_0,\omega_p$ and $\delta$ that determine the
characteristics of the gain medium. For example, suppose that
    \be
    \hbar\omega_0=5~{\rm eV},~~~~\hbar^2\omega_p^2=-0.04~{\rm eV}^2,~~~~
    \hbar\delta=1.25~{\rm eV},
    \label{fix1}
    \ee
Then $\omega_0,\omega_p$ and $\delta$, $\rho$ and $\sigma$ become
functions of $\Omega$ and $\omega$. For each choice of $\Omega$, we
can select a practically attainable range of frequencies $\omega$
and plot the parametric curve ${\cal C}^{\Omega}$ defined by
(\ref{e6}) in the $\rho$-$\sigma$ plane for this range of values of
$\omega$. The points of the intersection of ${\cal C}^{\Omega}$ and
$\tilde C^-_n$ correspond to the values of $\omega$ for which $k^2$
is a spectral singularity provided that (\ref{gamma=}) is also
satisfied. The latter condition fixes the value of $\alpha$. The
$\omega$ and $\alpha$ values obtained in this way depend on
$\Omega$, the label $n$, and another discrete label $\ell_n$ that
counts the number of times ${\cal C}^{\Omega}$ intersects $\tilde
C^-_n$ as one decreases the value of $\rho$ starting from its
largest allowed value, namely 1.

\begin{figure}
\begin{center} 
\includegraphics[scale=.85,clip]{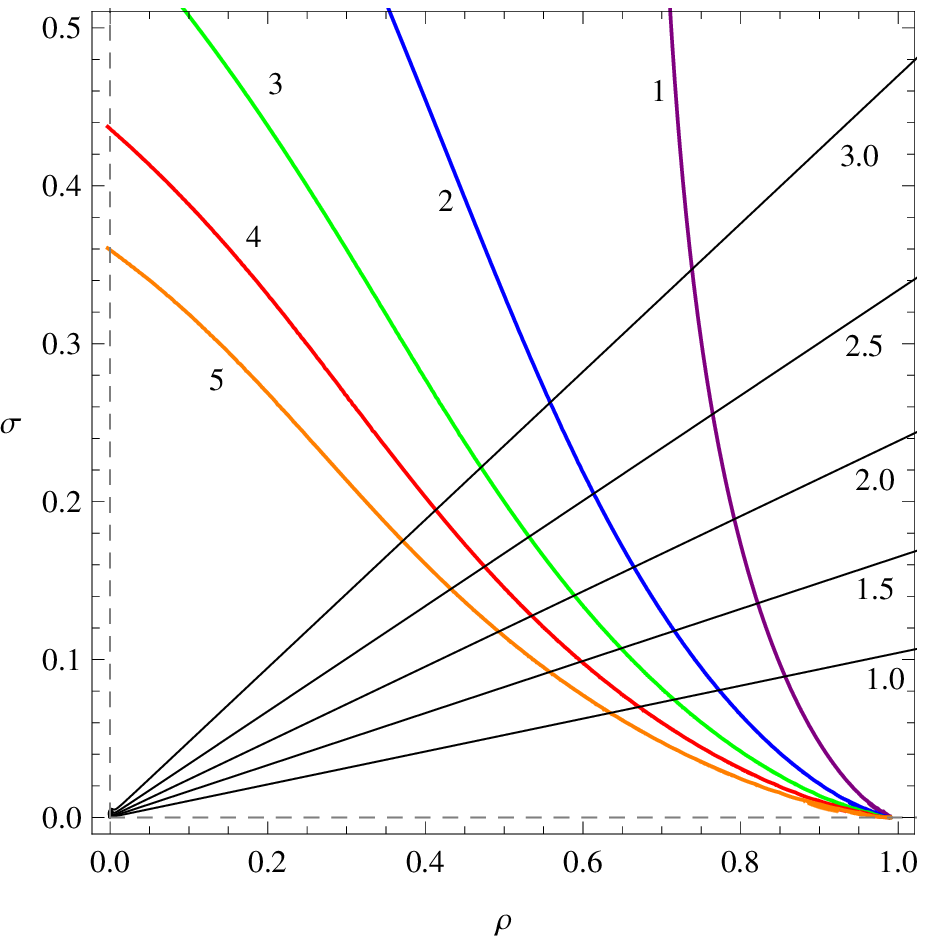} \vspace{.5cm} \\
\includegraphics[scale=.85,clip]{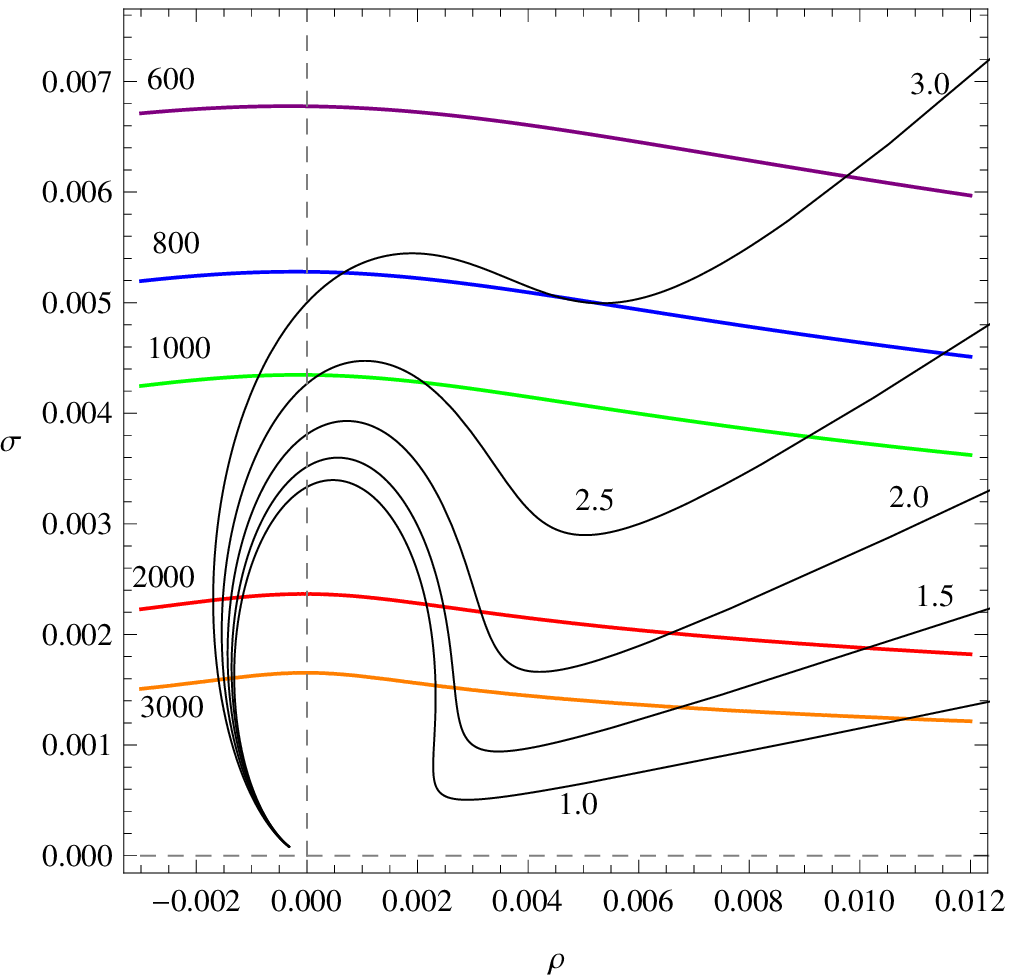}
{\caption{(Color online) Graphs of ${\cal C}^{\Omega}$ (thin black
curves) and $\tilde C_n^-$ (thick colored curves) for
$\omega_0,\omega_p,\delta$ as given by (\ref{fix1}) and $\Omega=1.0,
2.0, 2.5, 3.0~{\rm eV}/\hbar$ that respectively correspond to
$\beta/m=310.0, 155.0, 124.0, 103.3~{\rm nm}$. The top figure, shows
the intersection points of ${\cal C}^{\Omega}$ and the $\tilde
C_n^-$ with $n=1,2,3,4,5$. These correspond to $\ell_n=1$. The
bottom figure shows the intersection points of ${\cal C}^{\Omega}$
and $\tilde C_n^-$ with $n=600, 800, 1000, 2000, 3000$. For these
points $\ell_n=1,2$ or $3$. \label{fig4}}}
\end{center}
\end{figure}

To determine the precise values of $\omega$ and $\alpha$ for a given
choice of $\Omega$ and $n$, we plot the curves ${\cal C}^{\Omega}$
and $\tilde C_n^-$ in the $\rho$-$\sigma$ plane as in
Figure~\ref{fig4} and select an intersection point, i.e., fix
$\ell_n$.
We then determine the coordinates $(\rho_\star,\sigma_\star)$ of
this point graphically \footnote{Using Mathematica and without
making any special efforts we could zoom in the neighborhood of the
relevant intersection point and obtain $\rho_\star$ and
$\sigma_\star$ up to six significant digits for various values of
$\Omega$, $n$, and $\ell_n$.}. Next, we numerically solve for
$\omega$ in
    \be
    \rho=\rho_\star,~~~~\sigma=\sigma_\star,
    \label{numer}
    \ee
where the left-hand side of these equations are given by (\ref{e6}).
A consistency check on this calculation is that both of these
equations must give the same value for $\omega$. Having obtained
$\omega$, we compute $k$ using (\ref{e4a}), substitute this value of
$k$ and (\ref{numer}) in (\ref{gamma=}), and solve for $\alpha$ in
the resulting equation.

Figure~\ref{fig4} shows the plots of the curves ${\cal C}^{\Omega}$
and $\tilde C_n^-$ for various values of $\Omega$ and $n$. For every
$\Omega$, the curve ${\cal C}^{\Omega}$ intersects each of the
curves $\tilde C_n^-$ at least once and at most three times, i.e.,
$1\leq\ell_n\leq 3$. As one adopts sufficiently large values for
$\beta/m$ (sufficiently small values for $\Omega$), the number of
intersection points increases. This number equals to the number of
different frequencies $\omega$ at which a spectral singularity
occurs. Note, however, that each of these spectral singularities
correspond to a different value of the parameter $\alpha$.
Furthermore, because for each value of $\Omega$, ${\cal C}^{\Omega}$
intersects $\tilde C_n^-$ for all $n$, for every value of $\beta/m$
there is an infinity of choices for the $\omega$ (and $\alpha$)
values that yield a spectral singularity.

It turns out that for each value of $\beta/m$, one can select $n$ so
that the related spectral singularities correspond to experimentally
preferable values for $\alpha$ and $\omega$. For example, for the
principal mode $(m=1)$ of a waveguide of height $2\beta=1$~cm,
choosing $n=10000$ and $\ell_n=2$, we find a spectral singularity of
frequency $\omega\approx 2.1555$~ev$/\hbar$ (wave length:
$\lambda\approx 575.20$~nm) provided that we maintain an active
region of length $2\alpha\approx 2.8786$~mm. For future reference we
will use $\bigstar$ to label this spectral singularity. In fact, as
we would expect, changing the value of $n$ starting from such a
large initial value, we find a large number of closely spaced
spectral singularities all belonging to the experimentally
attainable range of values of $\alpha$ and $\omega$.

\begin{figure}[t]
\begin{center} 
\includegraphics[scale=.75,clip]{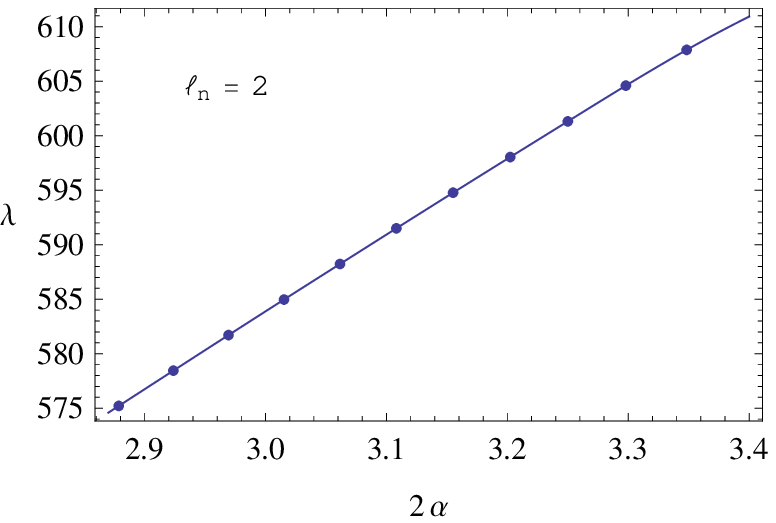} \vspace{.2cm} \\
\includegraphics[scale=.80,clip]{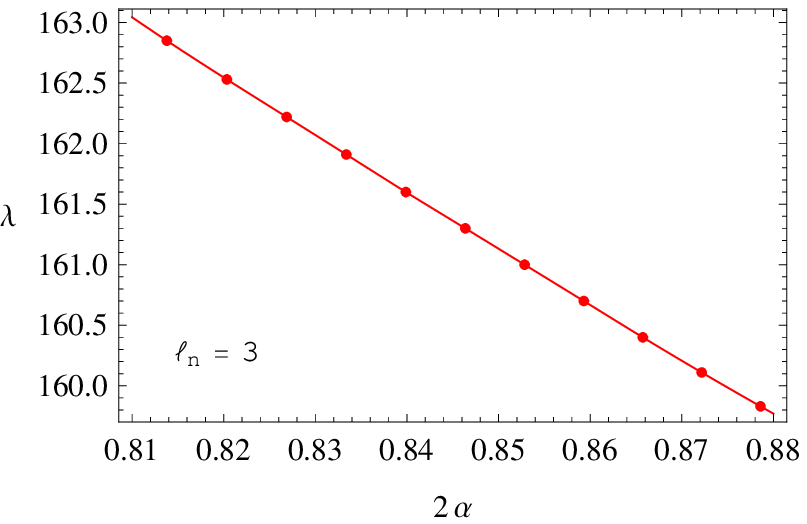}
{\caption{(Color online) Location of the spectral singularities in
the $2\alpha$-$\lambda$ plane for $2\beta/m=1$~cm and $10000\leq
n\leq 11000$. The numerical values of $2\alpha$ (length of the gain
region) and $\lambda$ (wavelength of the emitted wave) are given in
mm and nm, respectively. The displayed dots represent the spectral
singularities with $n=10000, 10100, 10200,\cdots,11000$ from the
left to the right. The curve segments joining adjacent dots
represent 99 isolated points each corresponding to a different
spectral singularity. The top and bottom figures show the spectral
singularities with $\ell_n=2$ and $\ell_n=3$, respectively. The
spectral singularity $\bigstar$ is also displayed. \label{fig5}}}
\end{center}
\end{figure}

Figure~\ref{fig5} shows the location of the spectral singularities
for $2\beta/m=1$~cm that we obtained for $10000\leq n\leq 11000$ and
$\ell_n=2,3$. For these choices of $n$, the spectral singularities
associated with $\ell_n=1$ correspond to extremely large
(experimentally unattainable) values for $2\alpha$.
As suggested by Figure~\ref{fig5}, using a waveguide of height
$2\beta=1$~cm and adjusting the length $2\alpha$ of the gain region
(in the millimeter range), we can generate spectral singularities at
almost every point of the visible spectrum or outside it. This is
also supported by a detailed examination of other choices for the
values of $2\beta$ and $n$ that we do not fully report here, except
for a few illustrative cases that we include in Tables~\ref{table1}
and~\ref{table2}.
  \begin{table}
    \begin{center}\vspace{.3cm}
    {\small\begin{tabular}{|c|}
    \hline
    $2\beta/m=1.0000~\mu$m\\
    {\begin{tabular}{|c|c|c|c|}
    \hline
  $\ell_n$ & $\lambda$ & $2\alpha$ & $\sqrt\varepsilon$\\
  \hline
  ~1~  & 1.6798 $\mu$m & 15.517 mm
  & $0.99919-6.1408 \times 10^{-5} i$ \\
  \hline
  2  & 614.03 nm & 3.2291 mm
  & $0.99910-2.1814 \times 10^{-4} i$ \\
  \hline
  3  & 162.61 nm & 0.81531 mm
  & $1.00045-2.6078 \times 10^{-4} i$
    \end{tabular}}\\
    \hline \hline
    $2\beta/m=1.0000$~mm\\
    {\begin{tabular}{|c|c|c|c|}
    \hline
    $\ell_n$ & $\lambda$ & $2\alpha$ & $\sqrt\varepsilon$\\
  \hline
  ~1~  & 1.9974 mm & 307.845 m
  & $0.99920-4.9699 \times 10^{-8} i$ \\
  \hline
  2  & 575.20 nm & 2.8786 mm
  & $0.99908-2.4333 \times 10^{-4} i$ \\
  \hline
  3  & 162.85 nm & 0.81379 mm
  & $1.00045-2.6259 \times 10^{-4} i$
    \end{tabular}}\\
    \hline\hline
    $2\beta/m=1.0000$~cm\\
    {\begin{tabular}{|c|c|c|c|}
    \hline
    $\ell_n$ & $\lambda$ & $2\alpha$ & $\sqrt\varepsilon$\\
  \hline
  ~1~  & 1.9982 cm & 6.7894 km
  & $0.99920-4.968 \times 10^{-9} i$ \\
  \hline
  \textbf{2} & \textbf{575.20} nm & \textbf{2.8786} mm
  & $\mathbf{0.99908-2.4333 \times 10^{-4} i}$ \\
  \hline
  3  & 162.84 nm & 0.81379 mm
  & $1.00045-2.6259 \times 10^{-4} i$
    \end{tabular}}\\
    \hline
    \end{tabular}}
    \caption{The wave length $\lambda$ of the TE wave, and the length
    $2\alpha$ and complex refractive index $\sqrt\varepsilon$ of the gain
    region corresponding to a spectral singularity with $\omega_p,\omega_0,
    \delta$ given by (\ref{fix1}) and $n=10000$. The boldface figures
    are those of the spectral singularity $\bigstar$.
    \label{table1}}
    \end{center}
    \end{table}
    \begin{table}
    \begin{center}\vspace{.3cm}
    \begin{tabular}{|c|}
    \hline
    $\ell_n=2$\\
    {\small{\begin{tabular}{|c|c|c|c|}
    \hline
  $n$ &  $\lambda$ ({\rm nm})& $2\alpha$ (mm)& $\sqrt\varepsilon$\\
  \hline
  2000 & 306.59 & 0.30685
  & $0.99902-1.1437 \times 10^{-3} i$ \\
  \hline
  3000 & 347.47 & 0.52173
  & $0.99893-7.763 \times 10^{-4} i$\\
  \hline
  4000 & 382.28 & 0.76534
  & $0.99895-5.8934 \times 10^{-4} i$\\
  \hline
  5000 & 415.09 & 1.03877
  & $0.99897-4.757 \times 10^{-4} i$\\
  \hline
    \end{tabular}}}\\
  \hline \hline
    $\ell_n=3$\\
    {\small{\begin{tabular}{|c|c|c|c|}
    \hline
    $n$ &  $\lambda$ ({\rm nm})& $2\alpha$ (mm) & $\sqrt\varepsilon$\\
  \hline
  2000 & 220.78  & 0.22059
  & $1.00055-1.1701 \times 10^{-3} i$ \\
  \hline
  3000 & 203.54 & 0.30504
  & $1.00064-8.043 \times 10^{-4} i$\\
  \hline
  4000 & 193.10  & 0.38589
  & $1.00062-6.1592 \times 10^{-4} i$\\
  \hline
  5000 & 185.45  & 0.46327
  & $1.00059-5.006 \times 10^{-4} i$
    \end{tabular}}}\\
    \hline
    \end{tabular}
    \caption{The wave length $\lambda$ of the TE wave, and the length
    $2\alpha$ and complex refractive index $\sqrt\varepsilon$ of the
    gain region corresponding to a spectral singularity with
    $\omega_p,\omega_0,\delta$ given by (\ref{fix1}),
    $2\beta/m=1.0000~{\rm cm}$, $\ell_n=2,3$, and $n=2000,3000,4000,5000$.
    \label{table2}}
    \end{center}
    \end{table}

As shown in Figure~\ref{fig5} and Tables~\ref{table1} and
~\ref{table2}, for $\ell_n=2$ the wavelength of spectral singularity
is an increasing function of the length $2\alpha$, while for
$\ell_n=3$ it is a decreasing function of $2\alpha$. This shows the
availability of a wide range of values of parameters of the system
that lead to a spectral singularity.

According to Table~\ref{table1}, for $n=10000$ and $\ell_n=2$, both
of the choices $2\beta/m=1$~mm and $2\beta/m=1$~cm give rise to
spectral singularities with the same values for the wave length
$\lambda$, the length parameter $2\alpha$, and the complex
refractive index $\sqrt\varepsilon$. Therefore, to observe this
spectral singularity one does not need to fine tune the height of
the waveguide. This is also true for $n\gtrsim 2000$ and $\ell_n=3$.

\begin{figure}
\begin{center} 
    \begin{tabular}{c}
    \includegraphics[scale=.09,clip]{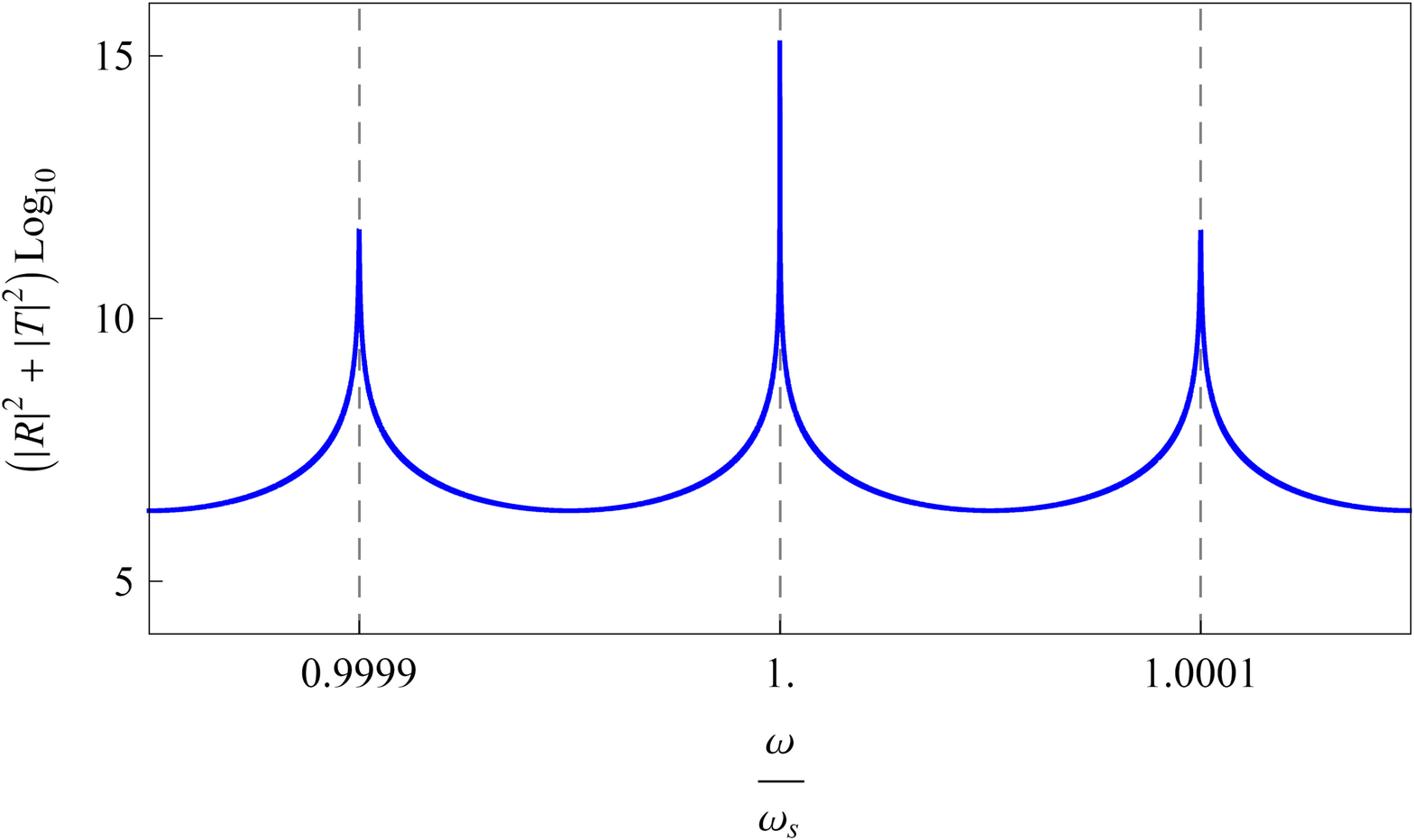}\\
    \includegraphics[scale=.09,clip]{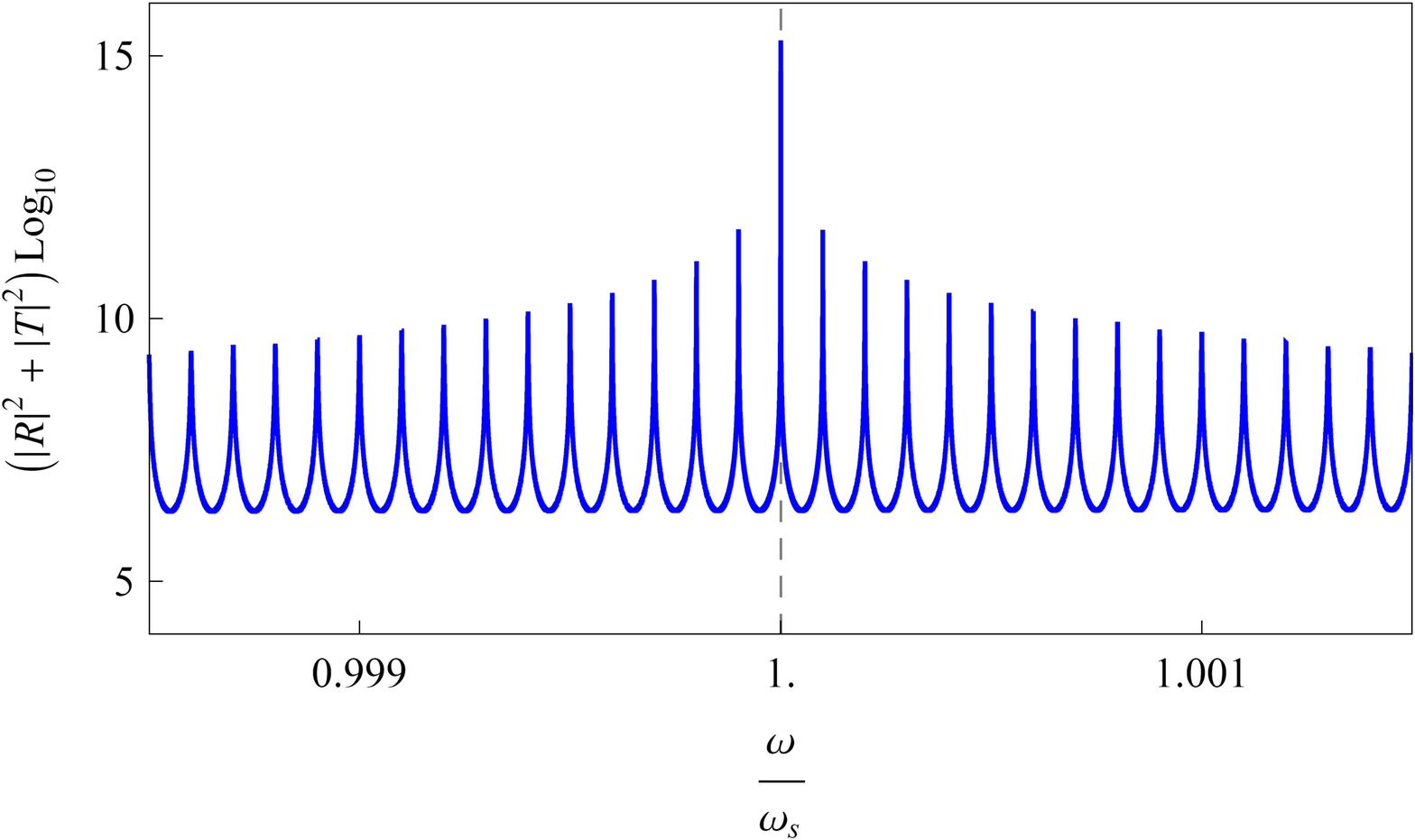}\\
    \includegraphics[scale=.09,clip]{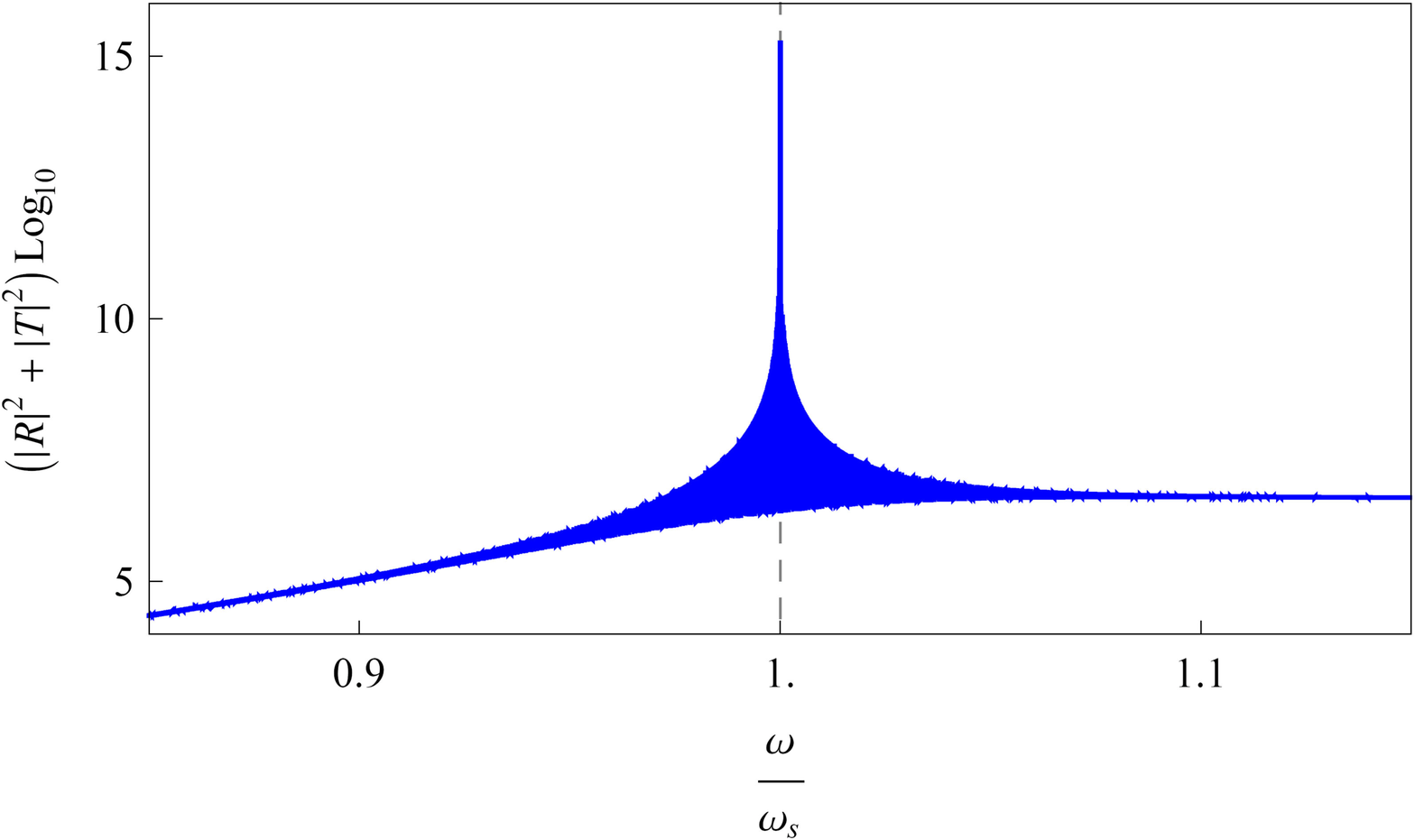}
\end{tabular}
{\caption{(Color online) Graphs of $\log_{10}(|T|^2+|R|^2)$ as a
function of $\omega/\omega_s$ for three different ranges of values
of $\omega/\omega_s$. $|R|^2$ and $|T|^2$ are the reflection and
transmission coefficients (See endnote~[27].) $\omega_s=2.15548~{\rm
ev}/\hbar$ is the frequency of the spectral singularity $\bigstar$
that occurs for $\omega_p,\omega_0,\delta$ given by (\ref{fix1}),
$n=10000$, $\ell_n=2$, $2\beta/m=1.00000$~cm, $2\alpha=2.878644$~mm,
$\lambda=575.2046$~nm, and $\sqrt\varepsilon=0.999081358-0.000243330
i$. The highest peak represents $\bigstar$. The other peaks
correspond to the spectral singularities with
$n=10001,9999,10002,9998,\cdots$. \label{fig6}}}
\end{center}
\end{figure}

Figure~\ref{fig6} shows the graphs of $\log_{10}(|T|^2+|R|^2)$ as a
function of $\omega/\omega_s$ where $\omega_s$ is the frequency of
the spectral singularity $\bigstar$ with $n=10000$ \footnote{Because
of the symmetry of the problem, and as we can easily verify by
inserting (\ref{Ms=}) in (\ref{amplitudes}), the left and right
reflection (and transmission) coefficients coincide; $R^r=R^l=:R$
and $T^r=T^l=:T$.}. This is represented by the peak at
$\omega/\omega_s=1$ that corresponds to an amplification of the
emitted electromagnetic energy by a factor of $\geq 10^{15}$. We can
increase this factor, if we use more precise values for the $\omega$
and $2\alpha$ parameters associated with $\bigstar$. The other peaks
represent the spectral singularities with
$n=10001,9999,10002,9998,\cdots$ and $\ell_n=2$. These are lower
than the central peak, because the corresponding spectral
singularities have a slightly different values for $\omega$ and
$2\alpha$ than those for $\bigstar$ that we have adopted in
preparing this figure. The location of the peaks can be easily
explained. Because the frequency of the spectral singularities with
$n\equiv 10000$ and $\ell_n=2$ decreases as a linear function of
$n$, changing $n$ by one unit, i.e., by a factor of $10^{-4}$, leads
to a change of $\omega$ by the same factor. This explains the
occurrence of the peaks at the $\omega/\omega_s$ values that differ
from unity by an integer multiple of $10^{-4}$. Another interesting
observation is that the amplification effect persists even if we
change $\omega$ by as much as $10\%$ (See the last graph depicted in
Figure~\ref{fig6}.) This is another indication of the suitability of
the proposed setup for an experimental realization of the spectral
singularity-related resonance effect.

\section{Conclusion}

Spectral singularities are the energies of a rather strange type of
scattering states that similarly to resonance states have infinite
reflection and transmission coefficients. Physically this means that
if one arranges to scatter a plane wave of energy close to that of a
spectral singularity, the system amplifies the wave enormously.
Therefore, spectral singularities are related to a novel resonance
phenomenon that awaits experimental verification.

For a real potential, the unitarity of dynamics implies that the
reflection and transmission coefficients add up to unity. Therefore,
spectral singularities cannot arise for a real potential. In this
article, we have explored the spectral singularities of a complex
barrier potential and showed how this potential enters the
description of the electromagnetic waves propagation in certain
waveguides. If we adjust the physical parameters of the waveguide
system so that the frequency of the incoming wave is close to that
of a spectral singularity, the waveguide emits a substantially
amplified wave of the same frequency from both ends. We expect this
phenomenon to find useful applications in optics, particularly
because we can adjust the resonance frequency to attain any desired
value.

A remarkable result of our investigation is that spectral
singularities of the complex barrier potential occur only for
certain values of the coupling constant with positive imaginary
part. While the mathematical reason for this phenomenon is obscure,
it has a simple physical justification. The condition that the
coupling constant has a positive imaginary part corresponds to the
case that the waveguide includes a gain region. This is necessary
for the existence of the resonance effect, because of the law of
conservation of energy.

One of the main goals of the present investigation was to examine
the range of critical values of the physical parameters of the
waveguide system that corresponded to the occurrence of spectral
singularities. We showed that indeed these parameters spanned a very
large spectrum of values. This makes the waveguide system we
considered in this article a plausible candidate for observing the
spectral singularity-related resonance effect.

\noindent{\textbf{Acknowledgments:}} I wish to thank Hossein
Mehri-Dehnavi and Ali Serpeng\"{u}zel for reading the first draft of
this article and making useful comments. I am also indebted to
Alphan Sennaro\u{g}lu and Aref Mostafazadeh for illuminating
discussions. This work has been supported by the Scientific and
Technological Research Council of Turkey (T\"UB\.{I}TAK) in the
framework of the project no: 108T009, and by the Turkish Academy of
Sciences (T\"UBA).

\end{document}